Tackling Challenges in Seebeck Coefficient Measurement of Ultra-High Resistance Samples with an AC Technique

Zhenyu Pan[1], Zheng Zhu[1], Jonathon Wilcox[2], Jeffrey J. Urban[3], Fan Yang[4], and Heng Wang[1]

1. Department of Mechanical, Materials, and Aerospace Engineering, Illinois Institute of Technology, Chicago, IL 60616, USA
2. Department of Chemical Engineering, Illinois Institute of Technology, Chicago, IL 60616, USA
3. The Molecular Foundry, Lawrence Berkeley National Laboratory, Berkeley, CA 94720, USA
4. Department of Mechanical Engineering, Stevens Institute of Technology, Hoboken, NJ 07030, USA

Abstract: Seebeck coefficient is a widely-studied semiconductor property. Conventional Seebeck coefficient measurements are based on DC voltage measurement. Normally this is performed on samples with low resistances below a few MΩ level. Meanwhile, certain semiconductors are highly intrinsic and resistive, many examples can be found in optical and photovoltaic materials. The hybrid halide perovskites that have gained extensive attention recently are a good example. Few credible studies exist on the Seebeck coefficient of, $CH_3NH_3PbI_3$, for example. We report here an AC technique based Seebeck coefficient measurement, which makes high quality voltage measurement on samples with resistances up to 100GΩ. This is achieved through a specifically designed setup to enhance sample isolation and reduce meter loading. As a demonstration, we performed Seebeck coefficient measurement of a $CH_3NH_3PbI_3$ thin film at dark and found $S$ = +550 µV/K. Such property of this material has not been successfully studied before.

1. Introduction

When a conductor is under a temperature gradient a voltage can be measured using a different conductor as probes. The measured voltage is proportional to the temperature difference at two contacts and the slope is the Seebeck coefficient $S$.[1] Seebeck coefficient is a key parameter for thermoelectric materials for solid state thermal-electrical energy conversion. It is also a fundamental material property of semiconductors, which, when combined with other properties, provides important information about electrical transport and electronic structures, such as the majority charge carrier type, carrier concentration, effective mass, band gap, *etc*[2] Knowing the Seebeck coefficient is of interest to a wide range of semiconductor research.

Measurement of Seebeck coefficient is essentially an open circuit voltage measurement plus a temperature measurement, which is generally considered easy. However, since the signal is a small voltage change usually less than 1 mV, the measurement becomes challenging when sample resistance is high. For instance, commercial test systems are usually rated for samples less than a few tens of MΩ.

There are many intrinsic semiconductors with very high resistivity, including organic semiconductors and large bandgap semiconductors for optical and optoelectronic applications. Hybrid halide perovskite semiconductors are a good example. Knowing Seebeck coefficients of these materials is more than a scientific challenge, because it provides useful information mentioned above about the free carriers. With sample resistances in the GΩ range, Seebeck coefficient measurements became a daunting task.

A strategy to reduce sample resistance is by changing sample dimensions. This strategy has limitations. Suitable sample length is needed for sufficiently uniform temperature difference. Increase of cross section faces instrument or sample preparation limitations. Moreover, change of sample dimension could also cause property change. As a result, it is not practical to reduce sample resistance over one order of magnitude through altering its dimensions.

The difficulty of measuring Seebeck voltage from high resistance samples comes from several sources. The first is circuit/meter isolation, the voltage-bearing wires as well as the meters need to be isolated with resistance much higher than the sample. The second is bias current of measuring instruments. Voltage measurements are open circuit but still need to draw charges (thus current) from the sample. For a 10GΩ resistor, even a 1pA of charge-drawing rate would cause 10mV voltage, which is fluctuating since the charge is not drawn at a steady rate. The Third is non-ideal sample behavior, real samples are not ideal resistors, any charge movement could cause random oscillations among local resistor-capacitor equivalent circuits, which almost never settle.

Most Seebeck coefficient measurements are DC method. Two metrologies are commonly used. The quickest method uses a temperature-sweep: while the temperature gradient is continuously increased/decreased, the voltage $V$ and temperature gradient $\Delta T$ are continuously and simultaneously recorded to calculate the slope. [3] [4] [5] This method usually completes a measurement in less than a minute. Alternatively, a steady-state method [6] [7] can be used, which first requires a stable temperature gradient across the sample to be established before $\Delta T$ and $V$ are recorded. Since it takes time to reach a steady state, and multiple $\Delta T$ are needed, the steady-state method takes longer (tens of minutes) for each measurement. Its advantage is that the voltage $V$ can be measured multiple times, which allows averaging over large numbers of readings, thus is necessary when the measurement is noisy, or when higher accuracy is needed. Generally, the steady-state method is suitable for a wider range of samples, and is used in commercial measurement systems such as ULVAC-ZEMs. Good practice for accurate Seebeck coefficient measurement using these methods has been discussed previously. [8] [9]

Taking average in steady state DC methods could compensate some of the fluctuations but will be less accurate as noise increases. Eventually, when the fluctuation amplitude is much greater than the signal, no reliable measurement can be performed. Previous development of measurement methods for high resistance samples focused on reducing DC voltage fluctuations, often by use of amplifiers with ultra-low bias current. [10, 11] This could offer small improvement but majority of the problems still remain.

It should be clarified that even though this problem seems to be due to 'high resistance' of the sample, the resistance itself is not the only issue, the non-ideal sample behavior associated with high resistance is probably more important. In fact, we had successfully measured Seebeck coefficient of samples >100GΩ using DC method with reasonable precision (±15%). [12]

In this report, we introduce an AC based measurement technique. By creating oscillating temperature gradients and read out the voltage response under same frequency using a lock-in amplifier, we isolate Seebeck signal out of excessive random noises. The result is clean voltage responses proportional to temperature difference, with negligible offset when $\Delta T$ is extrapolated to zero. Our method offers the ability to measure ultra-high resistance samples on the order of 100GΩ. This extends current Seebeck coefficient measurement capability by several orders of magnitude. It offers a tool to study high resistivity materials by electrical means.

2. Method

Lock-in amplifiers and phase-sensitive detection is often used to isolate signal of a given frequency in respond to a stimulus of that frequency, out of noises even several orders of magnitude greater than the signal, as long as they don't have the same frequency. Thus, if one could make the temperature gradient oscillates at a given frequency and measure the voltage signal under the same frequency using a lock-in amplifier, he could measure the Seebeck voltage with much reduced noise thus much improved accuracy. AC Seebeck measurement has been used previously to measure small Seebeck coefficients from metallic samples[13] [14]. With redesign of hardware and metrology, this technique can be applied to extend the Seebeck coefficient measurability in high resistance samples.

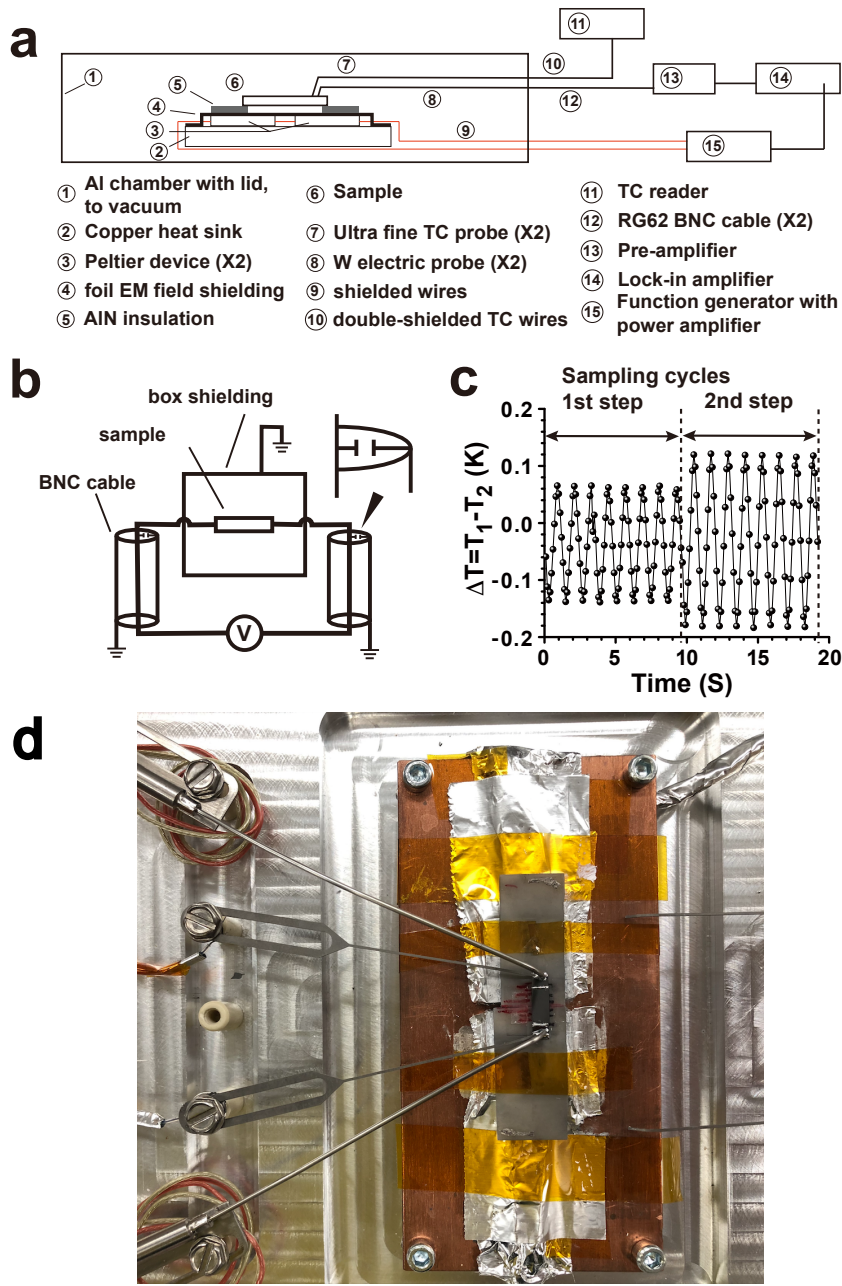

Figure 1. a) Schematic of the experimental setup. b) The equivalent circuit in Seebeck measurement highlighting cable shunt capacitance. c) An example of sinusoidal temperature difference oscillation at $f$=0.8Hz. d) A photograph of the sample area.

Fig. 1 shows a diagram and a picture of the test system. Temperature gradients are created along the horizontal direction by two Peltier devices. Foil shielding is applied over the Peltier devices to prevent electromagnetic interference caused by AC current flowing through these devices. The foil is grounded to chamber. Wires to these devices are also shielded with grounded Al foil. A 1mm AlN plate is put on top of the metal foil to keep sample electrically afloat for measurements.

Temperatures at both ends of the sample is measured by two ultra-fine, sheathed K type thermocouples with exposed tips. The stainless-steel sheath is 1mm in diameter (which is electrically afloat), tip size is about 1mm in diameter as well. The tips are carefully polished to create a flat contact surface with sample. We expect the thermocouples to have negligible thermal mass or cold-finger effect because of their size. A downward pressure is applied by springs at the far end of probes. Thermal grease is further applied to improve thermal contact. Voltages are measured by tungsten probes pressed against the sample in a similar way as the thermocouples. Voltage probes are separated from thermocouples. They are kept in isothermal regions to ensure accurate temperature reading. In-Ga-Sn eutectic or Ag paste were used to improve electric contact. The thermocouple wires outside of test chamber were connected to the TC reader using double-shielded thermocouple extension wires. Shielding is connected to chassis ground. The thermocouple reader is a Keithley 3706 test frame with a 3721 scanner card equipped with cold junction compensation.

The electrical leads outside of the test chamber are connected to a Stanford Research SR551 high impedance pre-amplifier with two 1ft long RG62 BNC cables. The choice of cable and its length is to minimize shunt capacitance. Power is supplied to the Peltier devices using a function generator through a power amplifier. The chassis of all instruments are connected to earth ground at a single point.

Measurements are performed in $N_2$ atmosphere or in air. Sinusoidal AC current of pre-decided frequency and amplitude was supplied. Time constant of the lock-in was set to be greater than 3× the oscillation period. After initial stabilization period, the voltage is recorded for three oscillation periods, then averaged to give final reading $V_{rms}$. After this the temperature difference is scanned, and the maximum and minimum values are recorded for each oscillation cycle, the averaged difference is used for $\Delta T_{p-p}$. Separating T reading from V reading process is necessary, as the scanning action of thermocouples causes changes to effective sample impedance, which compromises $V$ measurement. Three to five different oscillation amplitudes are used, the linear slope (which is always positive) $S_{rms}$ between $V_{rms}$ and $\Delta T_{p-p}$ is calculated. The magnitude of Seebeck coefficient is then calculated by $|S|=2\sqrt{2}\times S_{rms}$, the pre-factor reflects the ratio between RMS value and peak-to-peak value of a sinusoidal waveform.

### 3. Measurement examples
3.1 Validation with conductive samples

Fig. 2 shows the comparison of Seebeck coefficient of different types of samples with low to moderate resistances, measured with AC technique and DC technique. The AC and DC Seebeck coefficients are consistent in all cases. The voltage correction from the tungsten tips are applied. In the DC method, a temperature different within 10 K is applied; and in AC method, a temperature difference within 1 K is applied and the heating frequency is about 50 mHz.

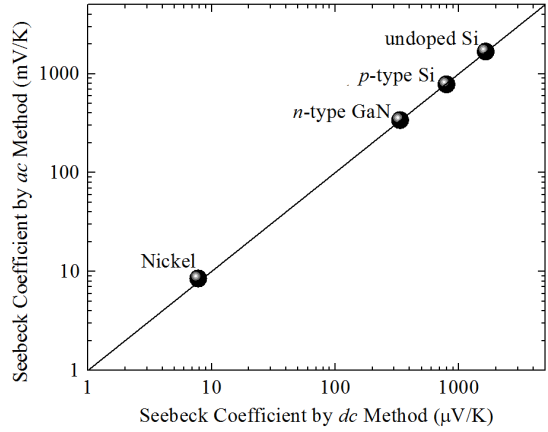

Figure 2. Seebeck coefficients of different samples with low to moderate resistances, measured with DC and AC technique

Fig. 3 shows the Seebeck coefficient of a p-Si (moderately doped) sample measured with AC technique using different heating frequencies. *I-V* curve between two voltage contacts measured a resistance of 1.3 kΩ. The top panel is the measured temperature oscillation amplitude using currents (peak value 30mA) of different frequencies. The lower panel shows an almost constant Seebeck coefficient measured at frequencies between 10mHz and 93mHz.

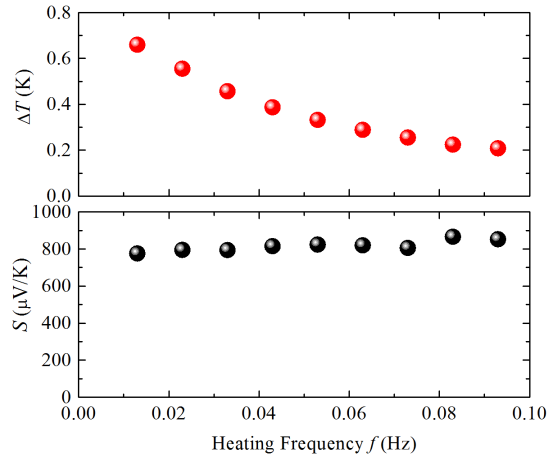

Figure 3. Seebeck coefficient of a p-Si sample measured with different frequencies.

3.2. High resistance samples

Seebeck measurements were performed on two high resistive samples. The first one is a piece of commercial semi-insulating GaAs single crystal. *I-V* test indicates a resistance about 1.6GΩ between voltage probes whose contacts are achieved by In-Ga-Sn eutectic compound. The *I-V* curve is linear up to +/- 6 V.

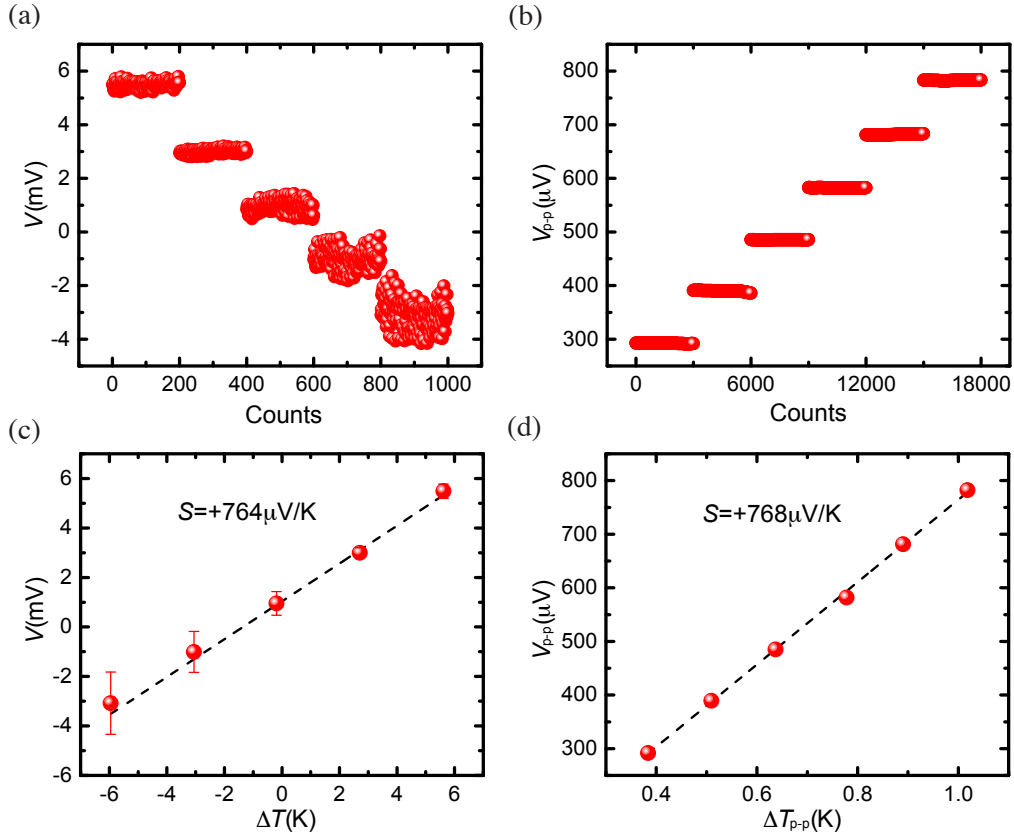

Figure 4. Seebeck coefficient measurement of a semi-insulating GaAs with DC and AC techniques. (a) Steady state voltage readings from DC method, (b) Voltage readings from AC method with frequency $f$ = 50mHz, (c) $V$-$\Delta T$ relation from DC method, due to the large error bars the Seebeck coefficient is subject to large uncertainty (d) $V$-$\Delta T$ relation from AC method, error bars for each data point are negligible.

Figure 4 shows the Seebeck coefficient measurement results from both DC and AC techniques. With DC technique, large voltage fluctuations up to 2mV were found under steady state, the voltage offset ($V$ at $\Delta T$=0) is around 1mV. As a result, even though Seebeck coefficient (764μV/K) can be determined from the slope of $V$-$\Delta T$, the uncertainty is quite large (+/-80μV/K). When AC technique is used, the voltage readout is almost flat with fluctuation only on the order of 3 μV. The resulting slope of $V$-$\Delta T$ is very linear with negligible uncertainty. Comparing to the 1mV DC voltage offset, the AC technique also removed the voltage offset such that only a negligible -4μV offset is found $\Delta T$ is extrapolated to zero.

A low frequency of 50 mHz was used in the comparison above. With higher frequencies, the 'Seebeck coefficient' will decrease with increased frequency as shown in Fig. 5. The reason for such dependence is the capacitive loading effect, which will be discussed.

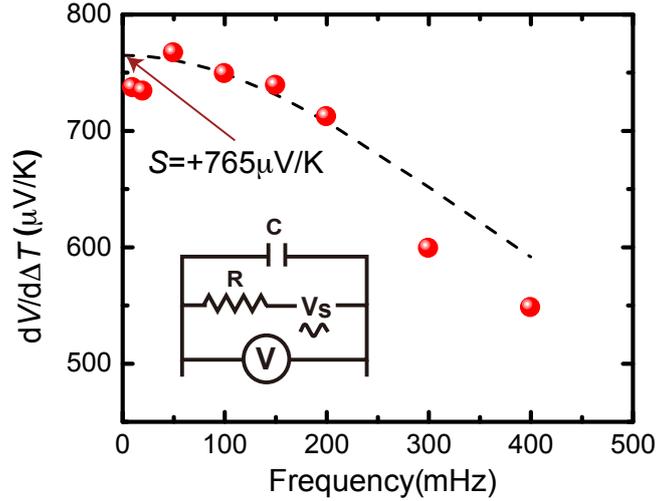

Figure 5. Frequency dependent d$V$/d$\Delta T$ results of GaAs. Inset: the equivalent circuit in ac Seebeck measurement.

The equivalent voltage measurement circuit considering cable shunt capacitance is shown in the inset of Fig. 5. This is a standard RC low-pass filter. If the effective capacitance $C$ and resistance $R$ are both known the signal attenuation can be simply calculated (see discussion). To determine $C$, we pass a AC current through a standard 1MΩ resistor and measure the voltage across it. From the frequency dependence of measured V, we calculated $C$ = 200 pF. Combining with $R$ = 1.6 GΩ the calculated frequency dependence of d$V$/d$\Delta T$ (the apparent Seebeck coefficient) matched with experimental result especially in the low frequency range, as shown in Fig. 5. By choosing sufficiently low frequencies, one can directly measure the Seebeck coefficient of a high resistance sample, however it will take very long time. On the other hand, higher frequencies can be used and true Seebeck coefficient can be derived from measured values. This reduces measurement time but will require good knowledge about the sample and test setup.

The second high-resistance sample is a $CH_3NH_3PbI_3$ thin film spin-coated on 1cm$^2$ borosilicate glass substrates under $N_2$ atmosphere. The synthesis is based on literature report [15]: 1 mole of Pb(Ac)$_2$·3H$_2$O plus 3 moles of $CH_3NH_3I$ were dissolved in 1L of dimethylformamide (DMF). Fresh solutions were used for spin coating at 3200 rpm for 40 seconds, followed by annealing at 70 °C for 2 minutes then 100 °C for 10 minutes. The film obtained have black, mirror-like appearance. Photoluminescence was measured using a Raman microscope with 532 nm laser excitation, and the emission peak was found at 766 nm as shown in Fig. 6a. Fig. 6b shows transmittance absorption spectrum measured with a UV-Vis-NIR spectrometer. The absorption edge is found at 810nm. $CH_3NH_3PbI_3$ is an important photovoltaic material, which is a highly intrinsic semiconductor. [16] [12] [17] No report can be found on successful measurement of Seebeck coefficient from a thin film. To measure this sample, In-Ga-Sn eutectic was used to make good contact. Ohmic *I-V* behavior was confirmed for currents up to ±9 pA and from the slope the resistance was determined to be 156 GΩ as shown in Fig. 7a. For this sample, DC method can no longer make acceptable measurement. The voltage offset and fluctuation completely overwhelmed Seebeck voltage which could be seen in Fig. 7b. AC method is the only option. Nonetheless, based on the sample resistance and shunt capacitance of the setup, it can be estimated that in order to obtain $V$-$\Delta T$ reflecting > 95% of true Seebeck coefficient, the frequency can't exceed 10 mHz. At this ultra-low frequency, the measurement will take over 20 hours, also a temperature oscillation is hard to be perfectly periodical thus even the lock-in reading is often found with fluctuation and $V$-$\Delta T$ not perfectly linear as shown in Fig. 7. In addition, a finite offset at zero $\Delta T$ is seen. We believe such offset is due to inaccurate temperature values (the AC methods in principle excluded any voltage offset), and the fact we see such offset only at ultra-low frequencies suggest the real temperature fluctuation at these frequencies is slightly different from ideal sinusoidal as assumed.

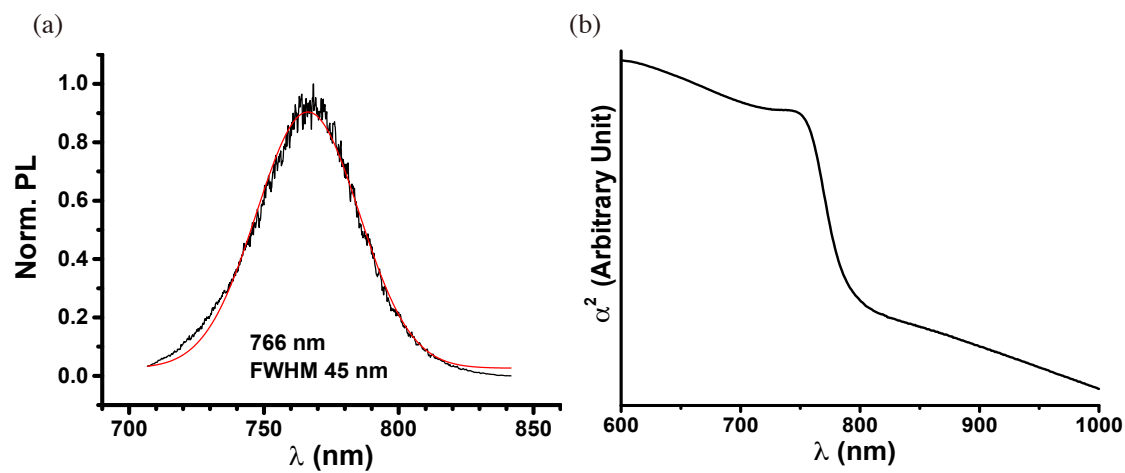

Figure 6. Basic characterization of the $CH_3NH_3PbI_3$ thin film (a) Photoluminescence peak at 766 nm with 532 nm excitation (b) Transmittance absorption spectrum indicating a band edge at 810 nm.

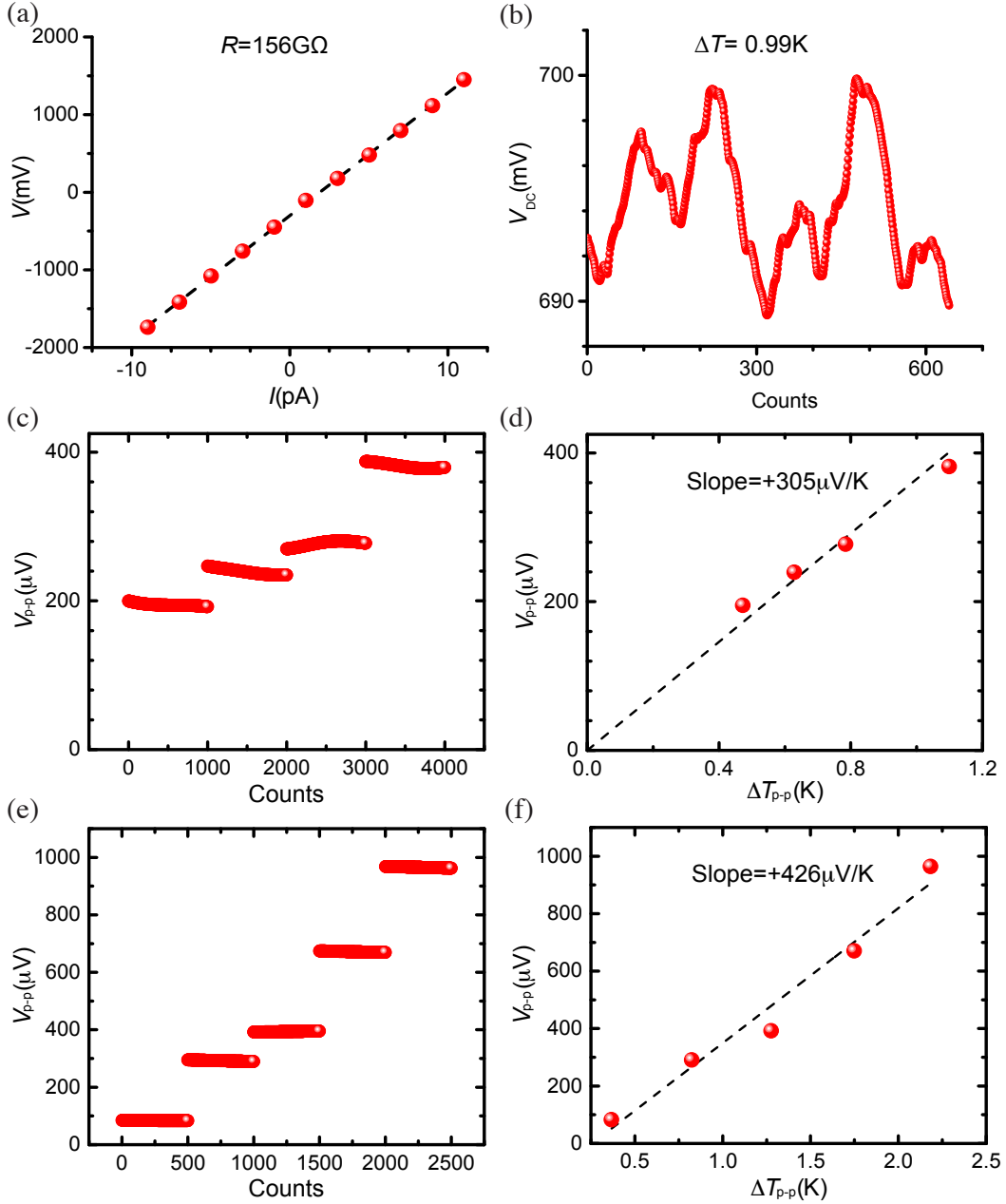

Figure 7. Measurement results on the CH$_3$NH$_3$PbI$_3$ thin film (a) *I-V* curve. (b) Voltage signal by dc method at $\Delta T$ = 0.99K. The voltage offset and fluctuation completely overwhelmed Seebeck voltage. (c)-(f) are voltage measurement by ac method: (c) Voltage readings over time with frequency *f* = 30mHz (only stabilized reading shown). (b) *V-$\Delta T$* relation at *f* = 30mHz. (e) Voltage readings overtime with *f* = 10mHz. (f) *V-$\Delta T$* relation at *f* = 10mHz.

We also studied the frequency dependence and interestingly, we found at relatively high frequencies the slope of *V-$\Delta T$* no longer decrease with *f* but instead became independent on *f*. The reason can be explained by a paradox: large resistors are not resistors. In analog circuitry, the small but finite parasitic capacitance in large resistors are not negligible, making them effectively low-pass *RC* circuits. For the CH$_3$NH$_3$PbI$_3$ perovskite film this is especially expected to happen [17]: Other than common features in highly intrinsic semiconductors, such as inhomogeneities, surfaces or grain boundaries, that could act as capacitors, the nature of perovskite structure, as well as the molecular dipole from $CH_3NH_3^+$ ion, both indicated stronger capacitive behavior. As a result, the $CH_3NH_3PbI_3$ perovskite film has an impedance that decreases with frequency as well. At (relatively) high frequencies, the resistive component is negligible, the voltage shunting ratio is determined by the parasitic capacitance of the sample compared to the shunt capacitance, which is a constant.

Figure 8 shows the frequency dependence of measured 'Seebeck coefficients', the equivalent circuit, and the calculated frequency dependence based on that circuit. *R* is the ohmic resistance of the sample 156GΩ, $C_1$ is the shunt capacitance of test setup which is 45pF (reflecting an upgrade after experiments on GaAs), $C_2$ is the parasitic capacitance of the sample. The observed frequency dependence can be explained with $C_2$ around 20pF. The extrapolated Seebeck coefficient is +550 µV/K. Alternatively, using ultra-low frequency oscillation of 5 mHz, the directly measured Seebeck coefficient was +500 µV/K, in reasonable agreement with the extrapolated result.

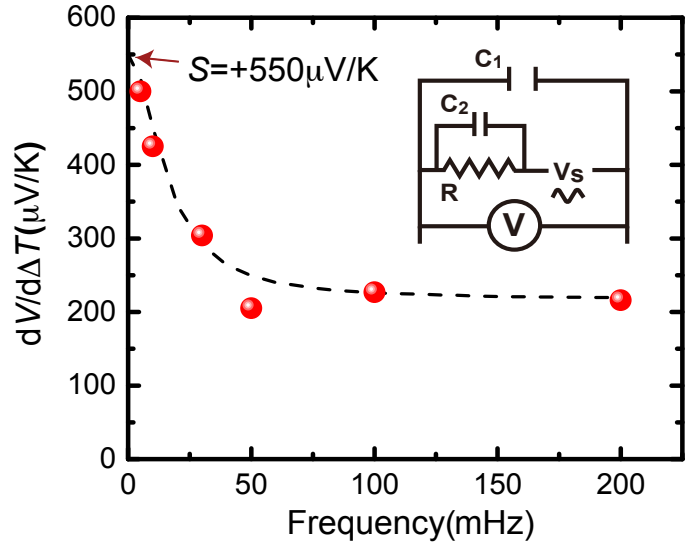

Fig. 8. Fitted frequency dependent d*V*/dΔ*T* from the $CH_3NH_3PbI_3$ thin film. Extrapolation indicates the DC Seebeck coefficient should be +550 µV/K. Inset: the equivalent circuit in ac Seebeck measurement.

4. Discussion
4.1. High resistance measurement considerations

AC Seebeck measurement has be developed to measure very small Seebeck coefficients from metallic samples. To use it for the other extreme of ultra-high resistance samples, there are specific considerations:

a) Means of temperature gradient generation. Two general ways have been used. First, one side of the sample can be radiated with a light source chopped at a certain frequency. This is used by multiple researchers studying metallic samples with small Seebeck coefficients. [13, 18] The advantage is that no electromagnetic interference can be introduced. Also, relatively high frequencies can be used, which could significantly reduce the time needed for the lock-in amplifier to reach a stabilized reading (which takes several tens of oscillation periods at these frequencies). Caution has to be given if this is applied to semiconductor samples. Photovoltaic effect may be present especially if the vicinity of voltage contacts is illuminated. Also, intrinsic semiconductors exhibit the photo-thermoelectric behavior so that the Seebeck coefficient changes under illumination. This can be avoided using a metal susceptor, but that requires metal deposition which complicates sample preparation. Second, resistive heaters can be used. [19] [20, 21] Oscillation frequency can be a few hundred mHz. At low frequencies, the cycle is on active heating but passive cooling, the asymmetric temperature waveform makes it not straightforward to determine the RMS value, which is used to calculate Seebeck coefficient as lock-ins read RMS values instead of peak. We overcome this problem in our design by using two Peltier devices connected in series with opposite polarities. A sinusoidal current (0.005 to 1Hz) is applied to both devices. The Peltier effect provided active heating and cooling linearly proportional to the current, resulting in a sinusoidal oscillation in temperature so the accurate RMS value can be determined. Note that the use of Peltier devices could induce electromagnetic interference at the same frequency as Seebeck voltage, which need to be prevented.

b) Contact and sample isolation. Ohmic contact is important for semiconductor measurements, whenever sourcing currents is needed. [21] [22] Seebeck measurements don't source current, and they measure the temperature coefficient of a potential difference. The requirement on contact is not as high, especially for DC measurements. When measuring a AC voltage, non-Ohmic contact could introduce rectifying effect leading to errors. Nonetheless, as long as the I-V relation is linear up to expected bias current and voltage signal, there should be no influence due to contacts. Better contact is still preferred though, as it reduces resistance making measurements easier. We have used In-Ga-Sn liquid metal, Ag paste, and direct mechanical contact for measurement on a semi-insulating GaAs, which showed sufficiently close results.

It is essential to keep the resistance between sample and ground much greater than the sample resistance. The sample should be afloat for voltage measurement (caution is needed as this potentially introduces electrostatic voltage dangerous for instruments), any unnecessary contact with the sample should be avoided. In conventional Seebeck measurement the voltage is read between the same type of wires of the two thermocouples. [23] [24] [25] However, thermocouple readers are not designed to have high input impedance, thus will essentially short the sample. Thus, when measuring high resistance samples the voltage probe and thermal couple probes have to be separated.

c) Determination of sign and magnitude of Seebeck coefficient
Unlike DC measurements, a lock-in amplifier does not tell the sign of the voltage response. The sign is determined by comparing the phase shift $\phi_v$ of measured voltage relative to reference signal (which is coupled to the current supplied to the Peltier devices), with the phase shift $\phi_T$ of the voltage of the thermocouple next to the $V_+$ voltage probe. Ideally $\phi_v$ and $\phi_T$ should either be equal or differ by 180°. In reality differences can be seen. Such differences are small so negative Seebeck coefficients are given by $\phi_v \cong \phi_T$ while positive ones are given by $\phi_v \cong \phi_T + 180°$.

Lock-in amplifiers read RMS values of voltage oscillation instead of peak values. On the other hand, thermocouple readers give real time $\Delta T$ values where it is easy to get peak-to-peak values. The magnitude of Seebeck coefficient is calculated by $|S|=2AV_{rms}/(\Delta T_{max}-\Delta T_{min})$, A is the crest factor which is the ratio of peak-to-peak value over RMS value for a given waveform. It is convenient to use sinusoidal or triangular temperature waveforms as their crest factor is well-defined.

  d) Circuit loading and frequency dependence

In most cases, the Seebeck coefficient measured with AC technique does not depend on frequency. Using higher frequencies is desired since it reduces the wait time to read from the lock-in. For example, $f = 21$Hz was used for metallic samples.[13] However, for high resistance samples, the measurement read out is frequency-dependent. To ensure a direct accurate measurement the highest usable frequency needs to be determined based on sample resistance and test setup.

Fig. 1 b) shows the equivalent circuit when an AC voltage across an ideal resistor is measured. In DC measurements, the input resistance of voltmeters need to be one to two orders of magnitude higher than the sample under test. Same requirement applies to AC measurements where resistance is replaced by impedance, which is made of both a resistive component and a capacitive component. Commercial lock-in amplifiers usually have input resistance of 10MΩ, which means it can't directly measure any sample with resistance above 1 MΩ. Pre-amplifiers need to be used to increase impedance to >1TΩ, making it possible to measure samples with resistance greater than 100GΩ. In addition to meter loading, the loading by capacitive component is also important, which could quickly compromise the measurement as the AC frequency $f$ increases. The impedance $Z$ of the $RC$ circuit shown in Fig. 5 is calculated by:

$$Z = \sqrt{R^2 + \overline{\frac{1}{2\pi f C}}^2} \qquad (1)$$

where $R$ is the sample resistance, $C$ is the shunt capacitance and f is the frequency. The capacitive component will cause the readout voltage $V_r$ to be only a portion of the source voltage $V_s$:

$$V_r = \frac{1}{2\pi f C Z} V_s \qquad (2)$$

For instance, for a resistive sample of 2GΩ, the $V_r$ will be compromised for $f > 0.15$Hz, when $V_r < 0.95\ V_s$, if the test setup has a capacitive component of 200 pF.

Capacitive component comes from both the test cable (shunt capacitance) and the pre-amplifier. The most commonly used BNC cable is RG-58 which has 85 pF/m capacitance. A reasonable estimate of capacitive input impedance of a pre-amplifier is 25pF. Hence it is common for a test setup to have 200pF capacitance. To minimize this,

RG-62 BNC cable can be used which has the lowest specific capacitance (47 pF/m) among commercial BNC cables. The length of cables should be kept at minimum by setting the amplifier close to test fixture.

Due to the shunt capacitance, direct measurement will eventually become impossible with AC method for high resistance samples, when $1/2\pi Cf$ becomes comparable to sample resistance $R$ even for the smallest $f$ (<5mHz). When $C = 45$pF, one can only measure samples up to 100GΩ, by setting the frequency $f = 10$mHz, in order to ensure $V_r > 0.95 V_s$. Fortunately for samples with higher resistances, fitting measurement values at different frequencies provides an indirect way to extrapolate true Seebeck coefficient values.

### 4.2. High Temporal resolution measurements

In addition to measuring ultra-high resistance samples, the exceptional noise rejection ratio from the AC method makes it possible to read out Seebeck voltage with minimum $\Delta T$ down to < 0.1 K. All Seebeck coefficient measurements need to create temperature differences across a sample, while the slope of $V$-$\Delta T$ is used to calculate $S$, the assumption is that change of $S(T)$ is negligible between $T-\Delta T$ and $T+\Delta T$. This is usually not a problem for most cases. However, if $S(T)$ has a strong temperature dependence (which for instance can be seen at the vicinity of phase transitions), $\Delta T$ of a few degrees could introduce unacceptable error. On the other hand, accurate, high temporal-resolution Seebeck coefficient through phase transitions could provide insights to changes in defects and electronic structure. Historically, AC technique has been used by different researchers to study the Seebeck coefficient of superconductor YBCO($YBa_2Cu_3O_{7-\delta}$) single crystals across its curie temperature[26][27][28][29]. Abrupt change and small peaks in Seebeck coefficient was observed and reflected by more than ten data points within a small temperature range less than five degrees.

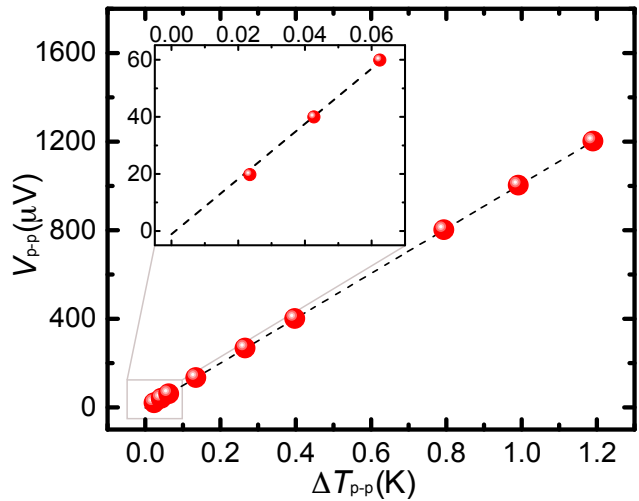

Figure 9. Seebeck measurement of a silicon sample, voltage is linear with $\Delta T$ down to 0.02K.

As another example, Figure. 9 shows the Seebeck coefficient of a piece of silicon sample (p-type) measured in this work using AC technique. With $\Delta T$ down to 0.02K, the measured $V$-$\Delta T$ still retains the same slope meaning the Seebeck coefficient can still be accurately measured. Combining this with high-resistance capability introduced here, we expect this AC method to be useful in studying the critical behavior of many different materials.

4. Conclusion

An AC technique for Seebeck coefficient measurement is developed here for samples with ultra-high resistances. Specially designed systems are needed for such measurements. In designing such a system a few factors need to be considered. First is the meter loading and shunt capacitance, both need to be minimized. The lock-in amplifier should be connected via a high impedance pre-amplifier to match the resistance of samples under test. The temperature measurement needs to separate from voltage probes as there is usually not high enough impedance with temperature measurement circuits. Second, the $RC$ settling behavior comes in even at very low frequencies when the resistance is beyond GΩ level, limiting the ability to perform direct measurements. Fitting can be employed to indirectly evaluated Seebeck coefficient with information of the system shunt capacitance and sample resistance. Measuring Seebeck coefficient from an ultra-high resistance sample is always challenging and each sample requires specific considerations. Nonetheless, we have demonstrated that such measurement is feasible on samples with resistances as high as 150GΩ.